\documentclass[prb,reprint,twocolumn,superscriptaddress,noshowpacs,notitlepage,longbibliography]{revtex4-1}

\usepackage{amsmath}
\usepackage{graphicx, epsfig}
\usepackage{verbatim}
\usepackage{color} 
\usepackage{subfigure}
\usepackage{url}
\usepackage{natbib}

\begin{document}
\title{Morphology Control of Epitaxial Monolayer Transition Metal Dichalcogenides}
\author{Akhil Rajan}
\email{ar289@st-andrews.ac.uk}
\affiliation{SUPA, School of Physics and Astronomy, University of St. Andrews, St. Andrews KY16 9SS, United Kingdom}
\affiliation{These authors contributed equally to this work}
\author{Kaycee Underwood}
\affiliation{SUPA, School of Physics and Astronomy, University of St. Andrews, St. Andrews KY16 9SS, United Kingdom}
\affiliation{These authors contributed equally to this work}
\author{Federico Mazzola}
\author{Philip D.C. King}
\email{philip.king@st-andrews.ac.uk}
\affiliation{SUPA, School of Physics and Astronomy, University of St. Andrews, St. Andrews KY16 9SS, United Kingdom}

\vspace{10pt}
\date{\today}
\vspace{20pt}

\begin{abstract}
To advance fundamental understanding, and ultimate application, of transition-metal dichalcogenide (TMD) monolayers, it is essential to develop capabilities for the synthesis of high-quality single-layer samples. Molecular beam epitaxy (MBE), a leading technique for the fabrication of the highest-quality epitaxial films of conventional semiconductors has, however, typically yielded only small grain sizes and sub-optimal morphologies when applied to the van der Waals growth of monolayer TMDs. Here, we present a systematic study on the influence of adatom mobility, growth rate, and metal:chalcogen flux on the growth of NbSe$_2$, VSe$_2$ and TiSe$_2$ using MBE. Through this, we identify the key drivers and influence of the adatom kinetics that control the epitaxial growth of TMDs, realising four distinct morphologies of the as-grown compounds. We use this to determine optimised growth conditions for the fabrication of high-quality monolayers, ultimately realising the largest grain sizes of monolayer TMDs that have been achieved to date via MBE growth.
 
\end{abstract}


\maketitle

\section{Introduction}

Transition metal dichalcogenides (TMDs), composed of a transition-metal (M) layer sandwiched between two chalcogen (X) layers, represent a particularly diverse materials family. In bulk, such covalently-bonded MX$_2$ monolayers are stacked with weak van der Waal's bonding between neighbouring layers. Depending on the filling of the transition-metal $d$-orbitals, a large variety of electronic properties are found, including semiconductors, metals, charge-density wave systems, superconductors, and topologically non-trivial materials.\cite{Chhowalla2013, wang2012electronics, xu2014spin, bahramy2018ubiquitous} Excitingly, their properties can be significantly modified by changing material's thickness down to the monolayer limit. Famous examples include a thickness-tuned cross-over from an indirect to a direct band gap in MoS$_2$~\cite{Mak2010, splendiani2010emerging}, the realisation of extremely high exciton binding energies as a result of reduced dielectric screening in the monolayer limit of various MX$_2$ semiconductors~\cite{raja2017coulomb}, and the emergence of a novel Ising superconductivity in single-layer NbSe$_2$, arising due to the combination of broken inversion symmetry and strong spin-orbit coupling.\cite{xi2016ising, bawden2016spin} 

The group IV and V TMDs, which are the focus of the current work, are perhaps most famous for the charge density wave (CDW) phases which they host. The group-V systems are $d^1$ metals, with large Fermi surfaces which undergo charge-ordering instabilities upon cooling.\cite{wilson1974charge} NbSe$_2$ and TaSe$_2$ additionally exhibit a superconducting instability at low temperature~\cite{Yokoya2518, kershaw1967preparation, revolinsky1965superconductivity}, while VSe$_2$ does not. The group IV system TiSe$_2$ also hosts a CDW-like phase~\cite{di1976electronic}, despite it being a very narrow-gap semiconductor~\cite{WatsonPRL}, and there has been substantial discussion over whether this compound may be considered as a rare realisation of an excitonic insulator.\cite{wilson1977concerning, cercellier2007evidence, kogar2017signatures} 

There has been substantial debate over how such interacting electronic states and phases evolve when thinned to a single monolayer. The charge-ordering temperatures have been reported to increase, decrease, or even vary non-monotonically with reducing sample thickness.\cite{doi:10.1063/1.4893027, Xi2015, Yu2015, Yoshidae1500606, Chen2015, Ugeda2015, doi:10.1021/acs.nanolett.8b01649} A robust ferromagnetic phase was reported to occur in the monolayer limit of VSe$_2$~\cite{Bonilla2018, xu2013ultrathin}, although several recent studies question this conclusion.\cite{doi:10.1021/acs.nanolett.8b01649, chen2018unique, fumega2019absence} To enable reaching a coherent understanding of the evolution of such quantum many-body states in monolayer TMDs, it is essential to develop improved methodologies for their materials growth. To this end, here, we report the fabrication of epitaxial monolayers of NbSe$_2$, VSe$_2$, and TiSe$_2$ using molecular-beam epitaxy (MBE). We investigate the effect of growth temperature, growth rate, and metal:chalcogen flux ratios on the uniformity and morphology of the monolayer films grown. We identify a key role of the transition-metal adatom mobility in dictating the growth dynamics, and through this develop strategies for the optimal growth of large area, high-quality epitaxial monolayers of transition-metal dichalcogenides.

\section{Results and discussion}
\begin{figure*}
\centering
\includegraphics[scale=.8]{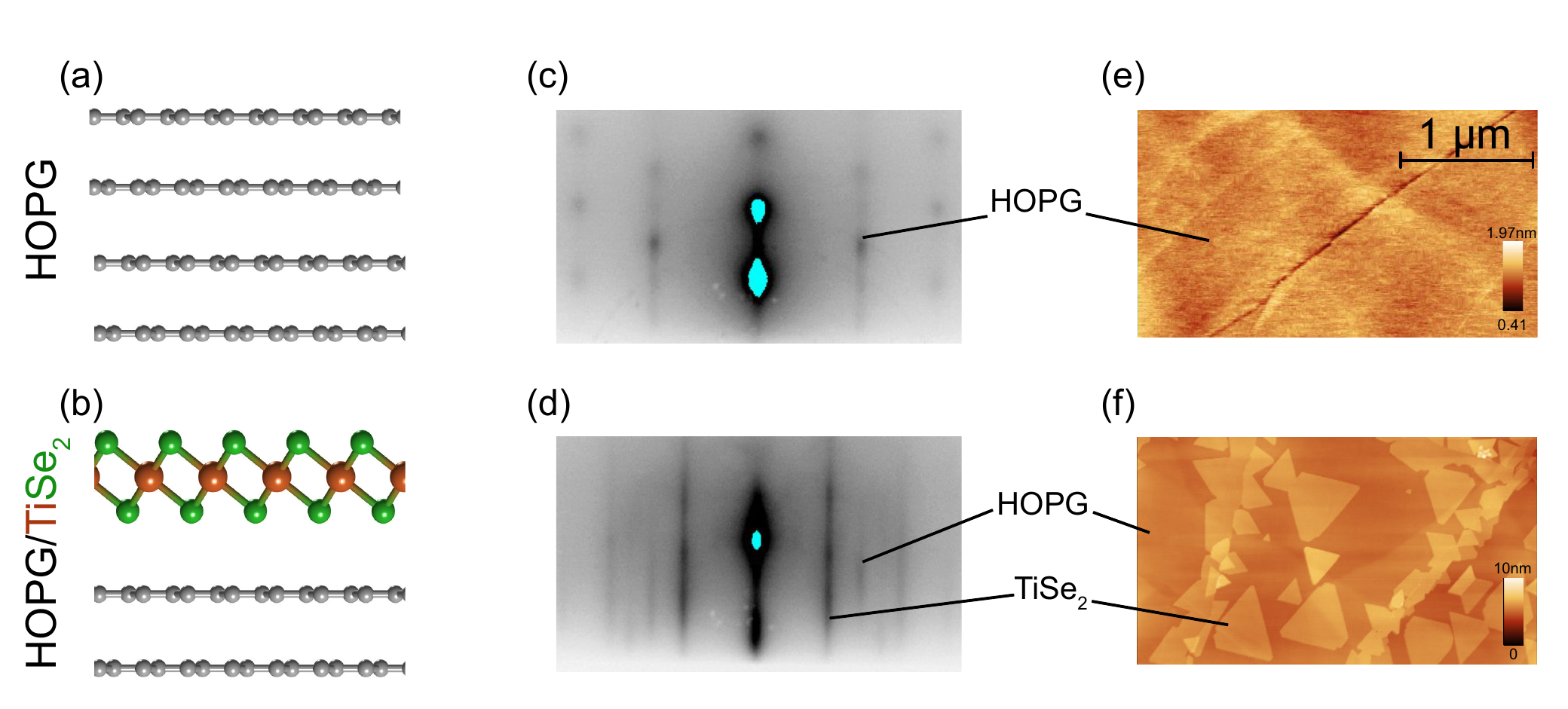}
\caption{Epitaxial growth of TMDs on HOPG. (a) Crystal structure of HOPG and (b) 1T phase of monolayer TiSe$_2$ grown on top of HOPG. RHEED showing (c) spotty streaks from the HOPG substrate before growth and (d) strong and streaky patterns following the growth of monolayer TiSe$_2$. AFM images showing the surface of a (e) bare HOPG substrate and (f) an as grown TiSe$_2$ monolayer surface.}
\label{Fig1}
\end{figure*}

Our approach is summarised in Figure~\ref{Fig1}. We employ highly-oriented pyrolytic graphite (HOPG) as a substrate throughout, which was cleaved immediately prior to loading into the growth system for each growth (see experimental section). This gives rise to a somewhat spotty (1$\times$1) RHEED pattern (Figure~\ref{Fig1}(c)). Atomic-force microscopy (AFM) measurements (Figure~\ref{Fig1}(e)) indicate a smooth surface, with occasional cleavage steps. Epitaxial TMD monolayers were grown on the cleaved substrate surface by co-evaporation of the transition-metal and chalcogen. The sticking coefficient of Se at the growth temperatures used (300-900 $^\circ$C) is very low compared to the metal species due to the huge differences in vapour pressures and the chemical environment. This necessitates a very high Se to metal flux ratio, which is also crucial in preventing the formation of 3D metal clusters via metal-metal bonding. A recent kinetic Monte Carlo simulation of the growth of WSe$_2$ has estimated the mean dwelling time of a Se adatom on the surface before desorption to be over four orders of magnitude less than that of the metal adatoms.\cite{Nie_2016} Similarly, the mean diffusion distance of Se as compared to the metal adatoms was two orders of magnitude shorter. During the growths performed for this work, we have therefore maintained a metal:Se flux ratio of at least 1:60, although for most parts of the study, a ratio as high as 1:500 was used. 

As the growth progresses, the RHEED pattern of the HOPG substrate begins to slowly fade, whilst a new pattern starts to appear which we attribute to the TMD epilayer. Towards the end of the growth, the new features become strong and streaky (Figure \ref{Fig1}(d)) confirming a flat morphology of the monolayer surface. From the spacing of the TMD RHEED streaks, we can extract a lattice constant for the TiSe$_2$ monolayer shown in Figure~\ref{Fig1}(d) of $3.52\pm0.05$~{\AA}. This is in excellent agreement with the in-plane bulk lattice constant of TiSe$_2$, despite the nearly 30 $\%$ lattice mismatch with the HOPG substrate. Equivalent results were obtained for VSe$_2$ and NbSe$_2$, confirming that the TMD monolayers are grown without strain and misfit dislocations, facilitated by a relaxed substrate-epilayer interaction at the interface via van-der-Waals epitaxy. Large-area AFM imaging (Figure~\ref{Fig1}(f)) indicates that growth yields a number of islands distributed across the sample surface. Around defects on the substrate (such as grain boundaries between neighbouring lateral domains with random in-plane rotational alignment, which are known to form in HOPG), there are a large number of nucleation sites and inhomogeneous growth is observed. Away from such substrate grain boundaries, there are a lower density of larger islands. In the following, we focus on the growth dynamics which dictate the morphology, size, and structure of these isolated islands, and elucidate the key parameters that can be tuned in order to optimise these.  
\begin{figure*}
\centering
\includegraphics[scale=.7]{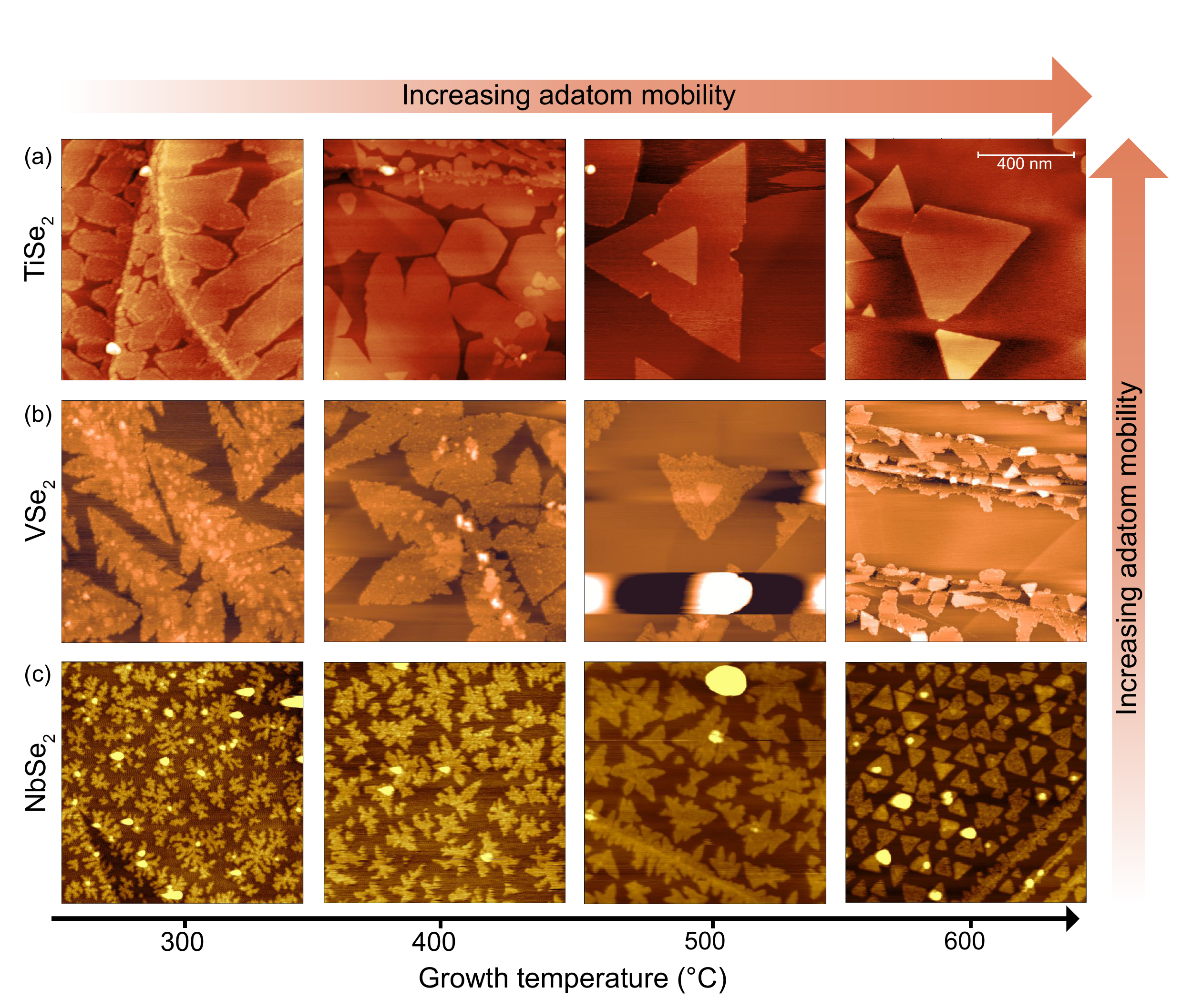}
\caption{Influence of adatom mobility on surface morphology. 1$\times$1~$\mu$m$^2$ AFM images showing surface morphology of three different materials, (a) TiSe$_2$, (b) VSe$_2$, and (c) NbSe$_2$ as a function of growth temperature. Both the metal and chalcogen adatom mobility increases from left to right, with an increasing growth temperature. Similarly, at a given growth temperature, the metal adatom mobility increases from bottom to top with increasing vapour pressure of the transition metal.}
\label{Fig2}
\end{figure*}

Figure~\ref{Fig2} shows AFM images of monolayer NbSe$_2$, VSe$_2$, and TiSe$_2$ grown on HOPG at growth temperatures of between 300$^\circ$C and 600$^\circ$C (throughout, we report the growth temperature as the temperature measured by a thermocouple positioned behind the sample plate). With an increasing growth temperature, the sticking coefficient decreases which results in a lower growth rate as evident from the smaller coverage of the epilayer (particularly clear for VSe$_2$). More importantly, changing both growth temperature and the transition metal atom leads to pronounced changes in the morphology of the as-grown monolayer islands. 

For NbSe$_2$, growth at the lowest temperature studied here leads to randomly-branched growth with very small feature sizes. We refer to this morphology as dendritic. With increasing growth temperature, a somewhat more symmetrical, but still branched, morphology of the growing islands is observed, while at the highest growth temperature ($T_g$) of $600^\circ$C, small triangular islands are formed, with a side length of ca. 50~nm. We note that at around $600^\circ$C growth temperature, NbSe$_2$ was previously reported to undergo a phase transition from the 1H (at low growth temperature) to the 1T (at higher growth temperature) polymorph, as judged from changes in the electronic structure measured using angle-resolved photoemission.\cite{MottNbSe2} Our own photoemission measurements (Supplementary Fig.~S1) indicate that for growths at 500$^\circ$C and below, our samples are purely in the 1H phase, whilst at a growth temperature of 600$^\circ$C, we still have predominantly 1H phase, but with a partial admixture of regions of 1T phase. No clear morphological differences are evident in different regions of the AFM scans shown in the bottom right panel of Figure~\ref{Fig2}, suggesting that the polytype does not have a major impact on the island morphology here, although this remains an interesting topic for future detailed exploration. 

VSe$_2$ and TiSe$_2$ are both expected to be stable in the 1T polymorph for all growth temperatures studied here. For both of these compounds, randomly branched growth is not observed at the lowest temperatures studied, in contrast to NbSe$_2$. Rather, at a growth temperature of 300$^\circ$C, a symmetrically branched growth mode is obtained. As is particularly clear for VSe$_2$, the growing islands have tree-like morphologies, with additional branching evident on the side of a growing spur, reminiscent of self-similarity. We thus attribute the symmetrically-branched structures as arising from a fractal growth mode. With increasing temperature, a trend towards a triangular growth mode is again observed. For a given growth temperature, the largest island sizes are observed for TiSe$_2$, with the smallest for NbSe$_2$.~\cite{Note1} The transition from branched to a triangular growth mode also occurs at lower growth temperatures for TiSe$_2$ vs. NbSe$_2$.

%~\footnote{The low coverage for VSe$_2$ at the highest growth temperatures here precludes us making a definitive statement at the highest growth temperatures, but generally VSe$_2$ lies between the case for NbSe$_2$ and TiSe$_2$.}.

The formation and evolution of these structures can be understood on the basis of varying adatom surface diffusion lengths. At a given growth temperature, the transition-metal mobility is the lowest for Nb atoms, while the Ti atoms are the most mobile. Moreover, higher growth temperatures lead to an increase in thermally promoted adatom surface diffusion, yielding longer surface diffusion lengths of both the metal and chalcogen species. The dendritic and fractal growth modes observed here can thus be understood due to the kinetic limitations of the adatoms at very low temperatures within a simple model of diffusion-limited aggregation.~\cite{PhysRevLett.47.1400} An adatom randomly diffuses on the substrate surface until it comes in contact with an already formed cluster or a nucleation site and sticks at the first point of contact. Once condensed at the edge of an island, edge diffusion is restricted or negligible at lower temperatures and this results in the formation of dendrites.\cite{PhysRevLett.67.3279} The dendritic growth observed for NbSe$_2$ at $T_g=300^\circ$C can thus be attributed to the extremely low mobility of Nb adatoms at lower temperatures. 

With increasing growth temperature, thermal excitation of the adatoms enables a moderate edge mobility. Randomly attached adatoms become more mobile and diffuse preferentially towards higher-symmetry bonding sites, enabling the steady coalescence of nucleating islands into morphologically more compact fractals. The mobility and directionality at this stage is still limited, however, and so the transition between the two growth morphologies is subtle. A key diagnostic is the increased symmetry of the fractal mode as compared to the dendritic one, similar to the morphological changes observed in the initial stages of growth of elemental metals on surfaces.\cite{Brune1994}  The transition is evident here with increasing growth temperature above 300$^\circ$C for NbSe$_2$. Dendrites are not, however, formed during VSe$_2$ or TiSe$_2$ growths even at $T_g=300^\circ$C, due to the relatively higher surface diffusion lengths of V and Ti adatoms as compared to Nb ones at that temperature.

The growth of compact triangular domains at higher temperatures differs from diffusion-limited aggregation. For the more compact growth, adatoms diffuse to an existing cluster, and then relax to a lower energy site through edge diffusion. As the rate of relaxation increases with respect to the rate of adatom diffusion to the cluster, a stoichiometric transition occurs from fractal growth to the more thermodynamically favourable triangular island growth mode. As the growth progresses, various islands begin to develop from different nucleation sites and they compete for the available adatoms. This naturally explains the steady transition from fractal to triangular domain growth mode evident for all three materials at intermediate/high growth temperatures discussed above. 

It is also evident, however, that there is a large difference in island size and density between the different compounds. This can again be understood from the varying adatom mobility of the different transition metals at a given growth temperature: In the absence of nucleation sites, an impinging atom diffuses until (within the surface dwell time) it comes in contact with another diffusing adatom which results in the formation of a seed. The mobility and stability of these seeds depend on the growth temperature. As the growth progresses, the number of seeds increases linearly until the the density is comparable to normal adatoms. At this point, island growth competes with any seed formation. At higher growth temperatures, the surface diffusion lengths of adatoms become larger than the mean island separation distances, which results in the adatoms diffusing into existing islands.\cite{BRUNE1998125} The significantly higher nucleation density present in NbSe$_2$ as compared to both VSe$_2$ and TiSe$_2$ can thus also be attributed to a lower thermally activated diffusion hopping rate of Nb vs. V or Ti adatoms at comparable temperatures.

\begin{figure*}
\centering
\includegraphics[scale=.7]{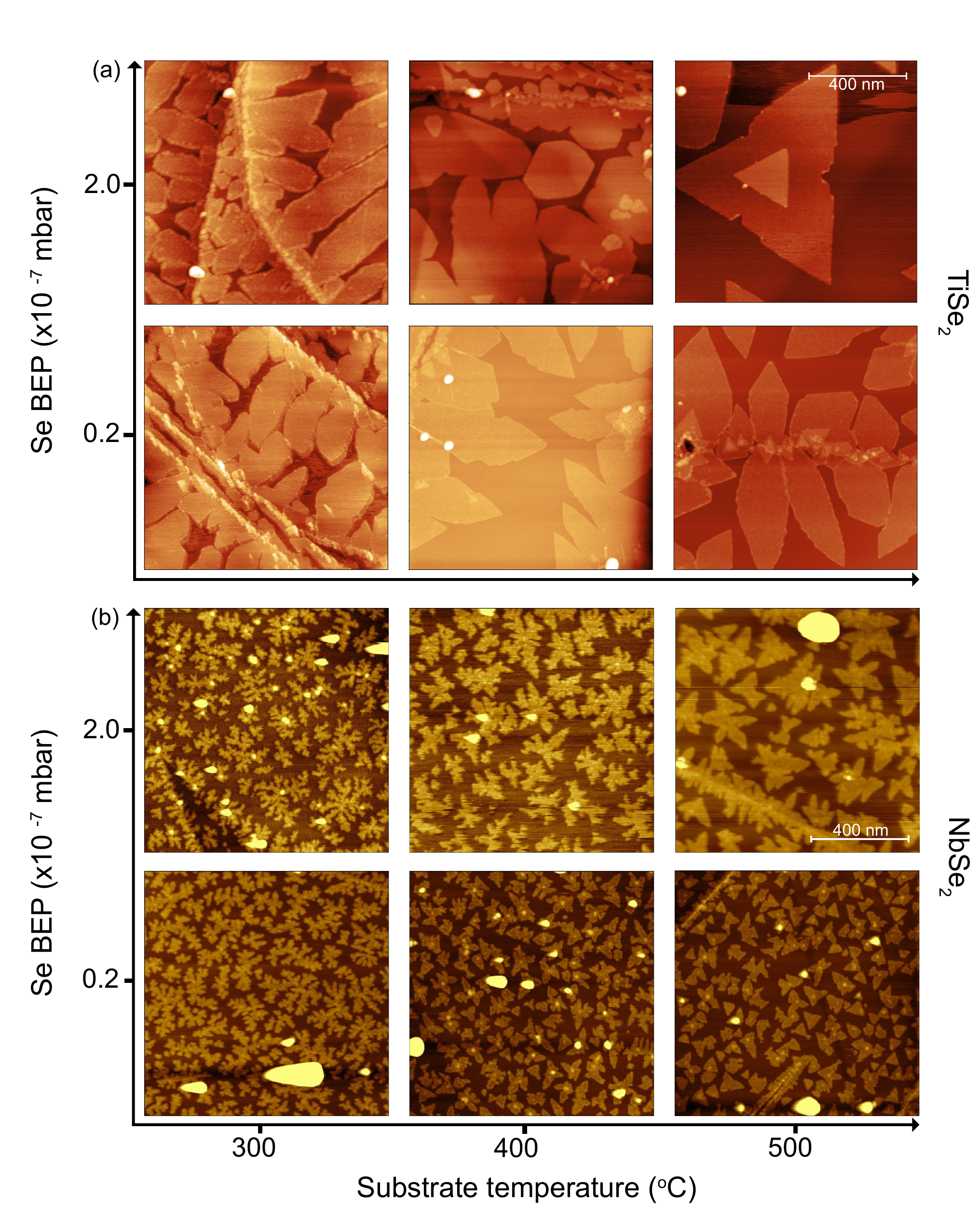}
\caption{Effect of Se flux. 1$\times$1~$\mu$m$^2$ AFM images showing the influence of increasing Se flux and growth temperature on the island morphology of (a) TiSe$_2$ and (b) NbSe$_2$ monolayers.}
\label{Fig3}
\end{figure*}

The above results demonstrate the major impact that variations in the adatom surface diffusion length, governed by changing growth temperature and transition-metal atom, have on the morphology of TMD monolayers grown by MBE. Nonetheless, other parameters can also influence the fractal to triangular domain transitions observed above. In the following, we focus on TiSe$_2$ and NbSe$_2$ as these show the extremes of behaviour of transition-metal surface diffusion. Figure \ref{Fig3} shows the morphology of TiSe$_2$ and NbSe$_2$ monolayers grown at temperatures between 300$^\circ$C and 500$^\circ$C under two different Se fluxes, corresponding to a Se beam equivalent pressure (BEP) of $\sim2\times10^{-8}$~mbar and $\sim2\times10^{-7}$~mbar, respectively. The AFM scans from the growth in the more Se-rich conditions is reproduced from Figure~\ref{Fig2} to aid comparison. 

\begin{figure*}
\centering
\includegraphics[scale=.7]{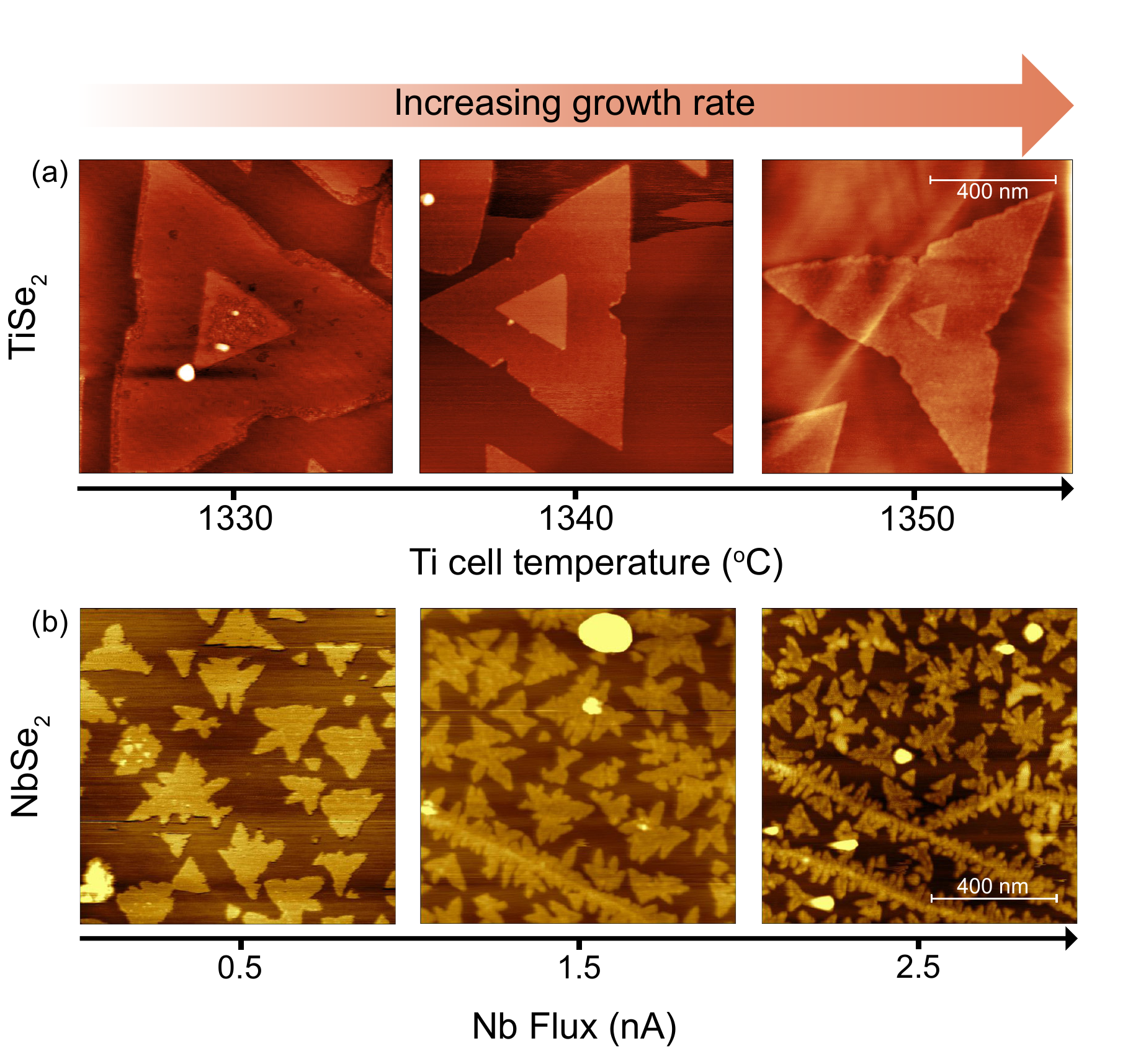}
\caption{Growth rate dependence. 1$\times$1~$\mu$m$^2$ AFM scans indicating the influence of growth rate on island morphology of (a) TiSe$_2$ and (b) NbSe$_2$ monolayers, as controlled by varying Ti effusion cell temperature and Nb flux from the e-beam source, respectively. The growth times are increased with decreasing growth rate, in order to obtain a similar surface coverage for all growths.}
\label{Fig4}
\end{figure*}

While qualitatively the same transitions from dendritic to fractal to triangular growth modes are still evident, this evolution is slowed down when growing using the lower Se flux. A noticeable change for the high Se, high temperature growth is the increased domain sizes as compared to growth in a lower Se flux. We stress that even at the lower Se BEP of $\sim2\times10^{-8}$~mbar, this still corresponds to a very Se-rich growth condition with a high metal:Se ratio of $\sim$~1:60. Nontheless, given the extremely volatile nature of Se, we find that further increase above this value leads to a significant increase in the effective sticking co-efficient, particularly at the highest growth temperatures. Considering the constant metal fluxes used here, it is evident that the excess Se impinging on the surface takes part in bonding with the metal adatoms and by means of edge diffusion forms the energetically favourable triangular domains. The increased surface diffusion lengths at higher temperature enables the formations of larger islands. It is also evident that in the absence of any excess Se, the extra metal adatoms otherwise available for bonding do not form any metallic clusters, possibly due to a combination of lower sticking coefficients and higher adatom mobilities. We also note that even at the lowest temperature we have used, the Se sticking coefficient has a huge dependence on the metal fluxes and hence Se atoms do not take part in the growth in the absence of the metal adatoms.

Given the pronounced influence of adatom mobility on the morphology of the synthesised epilayers outlined above, it is of interest to also investigate the influence of growth rate. In the following, we thus fix the growth temperature to 500$^\circ$C and the Se BEP to $\sim\!2\times10^{-7}$ mbar, and vary the impinging transition-metal flux. Figure~\ref{Fig4}(a) shows $1\times1$~ $\mu$m$^2$ AFM images of TiSe$_2$ samples for which the Ti effusion cell temperature was varied from 1330$^\circ$C to 1350$^\circ$C. We note that since the transition metal flux is very low (BEP $\sim\!6\times10^{-10}$~mbar), the change in Ti flux is modest for this change in cell temperature, and the error in measuring the BEP is high at these low values, we report simply the cell temperature used here.  To compensate the changes in surface coverage due to a varying metal flux, growth times were adjusted accordingly (from 55 minutes for the sample grown with 1350$^\circ$C cell temperature to 90 minutes for the sample growth with 1330$^\circ$C cell temperature) to maintain approximately equivalent coverage for the different growth rates used.

Two key features are evident in the case of TiSe$_2$ (Figure~\ref{Fig4}(a)). First, the triangular domains seen on all three AFM images consist of a larger monolayer island with side length varying from $\sim\!0.9 - 1.2$~$\mu$m and a smaller bilayer island on top. The monolayer island is not a perfect triangle, but shows some deformations along its edges. This is most pronounced for the fastest growth rate (highest Ti cell temperature), but similar deformations are evident even for the samples grown more slowly. The growth temperature here ($T_g=500^\circ$C) is at around the temperature at which the transition from fractal to triangular growth was found for TiSe$_2$ in Figure~\ref{Fig2}(a). Our growth-rate dependent studies here suggest that, at around this transition temperature, the growth of a monolayer is initiated by the formation of three fractal islands separated by 120$^\circ$ rotation. These fractal islands are originated from the same nucleation site and as the growth progresses, slowly evolve and merge to form a large triangular island. As seen from Figure~\ref{Fig4}(a), this transformation is highly growth rate dependent: at faster growth rates, there is not enough time for the adatoms to participate in edge diffusion, and the domains remain more fractal, while at slower growth rates, the adatoms have a longer time for edge diffusion, thus facilitating the formation of larger triangular grains. Slow growth rates are thus clearly preferable to generating large triangular islands.

These are not isolated monolayers, however, but have a small bilayer region forming at the middle of the island. The bilayer exhibits good epitaxial registry with the underlying monolayer. Interestingly, unlike the monolayer, we find that the bilayer region forms as a near-perfect triangle immediately from the initial stages of growth. This is likely due to the differences in growth kinetics when a layer is grown on a graphite substrate vs.\ a monolayer substrate of the same kind as the growing epilayer, which thus acts a favourable substrate. A reduction in size of the bilayer is evident with increasing growth rate. We speculate that this reflects the fact that, for the faster growth rates, the three 120$^\circ$ rotated fractal legs of the underlying monolayer have lateral dimensions less than the surface diffusion lengths of adatoms. Thus, adatoms which absorb on the monolayer surface can diffuse to the edge of the monolayer, where they can then participate in edge diffusion of the monolayer itself. Less adatoms thus contribute to forming a bilayer region. In contrast, for the slower growth rates, the monolayer becomes more triangular and its centre becomes further away from any nucleation edges, which in turn ultimately favours nucleation of a second layer atop the monolayer. This suggests that the formation of bilayer patches can be reduced by again increasing the surface diffusion length of the adsorbed adatoms, such that they reach the edge of the growing monolayer island within their surface diffusion time, and thus participate in edge diffusion, resulting in the formation of larger monolayers without bilayer growth. Consistent with this, we note that for TiSe$_2$ growth at a temperature of 600$^\circ$C (Figure~\ref{Fig2}), for which the adatom mobility is consequently increased as compared to the growths shown in Figure~\ref{Fig4}, no bilayer formation is observed. In fact, these monolayer islands, with edge length of ca. 600~nm, are -- to our knowledge -- the largest pure monolayers (i.e., without partial bi- or multi-layer coverage) of any TMDs achieved to date via MBE growth.

Figure~\ref{Fig4}(b) shows equivalent growth-rate dependent measurements for NbSe$_2$. Here, a larger change in flux from 0.5 to 2.5 nA (as measured by a flux monitor integrated into the electron-beam evaporator used for the evaporation of Nb) was used, with the corresponding growth times changed from 540 to 35 minutes, respectively, in order to maintain approximately equivalent surface coverage. As evident in Figure~\ref{Fig4}(b), such changes in the NbSe$_2$ growth rate have a significant influence on both the onset of nucleation and the sizes of islands. At faster growth rates, there is an increased number of nucleation sites and resulting islands. However, when the growth is slowed down, the nucleation site density decreases as the adatoms have more time to migrate over longer distances, increasing their probability of a subsequent encounter with an existing island. This enhancement in the surface migration length also gives rise to larger monolayer islands. As for TiSe$_2$, there is evidence of some fractal to triangular domain transformation occurring at the lowest growth rates. Nonetheless, there are no clear triangular domains formed for NbSe$_2$ here, as the 500$^\circ$C growth temperature used is still below the temperature where this transition occurs (Figure~\ref{Fig2}), due to the significantly lower adatom surface diffusion lengths of Nb atoms as compared to Ti. It is clear, however, from the measurements shown in Figure~\ref{Fig4} that the transition from a fractal to triangular growth mode is not simply a function of the surface adatom mobilities, but can also be strongly modified by the growth rate used, as well as the ratio of metal:chalcogen flux, as shown in Figure~\ref{Fig3}.

\begin{figure*}
\centering
\includegraphics[width=\textwidth]{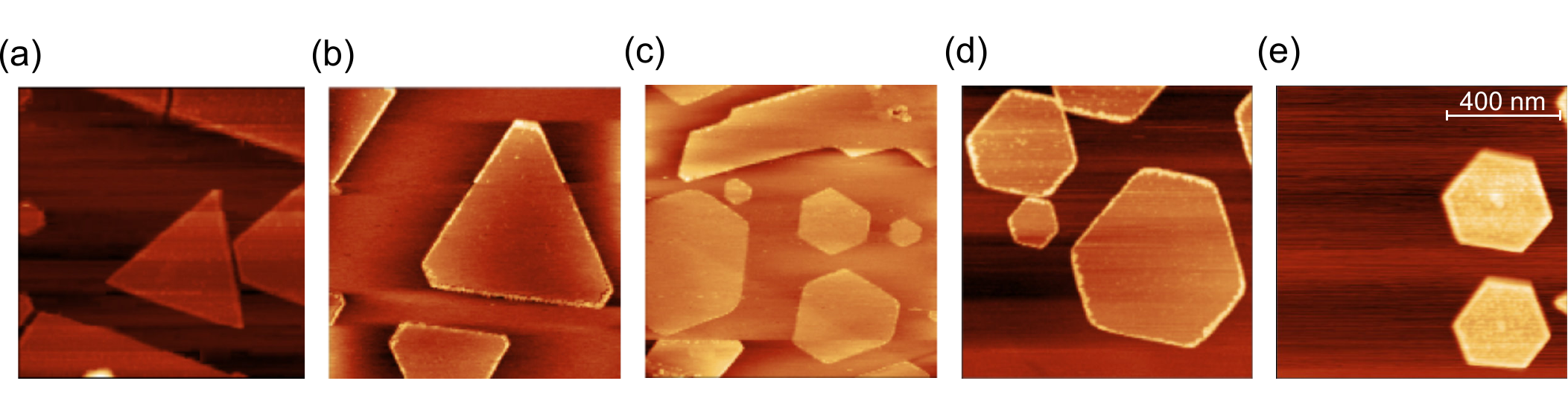}
\caption{Transition from triangular to hexagonal TiSe$_2$ islands. With increasing growth temperatures of (a) 600$^\circ$C, (b) 750$^\circ$C, (c) 800$^\circ$C, (d) 850$^\circ$C and (e) 900$^\circ$C, a gradual transition is observed, whereby the tips of the triangular island are truncated, transforming the island monolayer into a hexagonal shape. Samples (a-d) were grown using the same metal flux (Ti at 1340 $^\circ$C), whereas a lower metal flux (Ti at 1330 $^\circ$C) was used for sample (e). The growth times were adjusted accordingly to compensate for the slower growth rates at elevated growth temperature [(a): 150 minutes; (b): 300 minutes; (c-d): 350 minutes; (e) 450 minutes]. }
\label{Fig5}
\end{figure*} 

From the above, it is thus clear that enhancing adatom mobility and utilising slow growth rates are key to obtaining more compact and thermodynamically-favourable morphological configurations of the epitaxial TMD islands, and for realising true monolayer growth without additional bilayer patches. To explore this further, and to investigate whether other close-packed configurations may be obtained, we have synthesised TiSe$_2$ monolayers using even higher growth temperatures. Figure \ref{Fig5} shows AFM images of the resulting TiSe$_2$ monolayers grown at 600, 750, 800, 850 and 900 $^\circ$C. We find that the triangular domains discussed above slowly transform into hexagons, via a gradual truncation of the tips of the original triangular domain with increasing growth temperature. The process starts when increasingly energetic atoms attached to the three corners of a triangle undergo edge diffusion at elevated temperatures. In CVD synthesis of WSe$_2$, a transition from islands of triangular morphology to hexagonal islands was previously observed to be associated with a cross-over from monolayer to multi-layer structures.\cite{doi:10.1021/acsnano.5b01301} A transition to hexagonal monolayer patches was also reported during the CVD growth of MoS$_2$, where the change in morphology was attributed to the changes in the Mo:S ratio of the precursors.\cite{doi:10.1021/cm5025662} In contrast, the triangular to hexagonal transition observed in our work can be attributed simply to the increasing adatom mobilities with temperature, and thus reflects the intrinsic stability of the hexagonal morphology of the as-grown layer given high adatom diffusion lengths.

\section{Conclusions}
\begin{figure*}
\centering
\includegraphics[scale=.6]{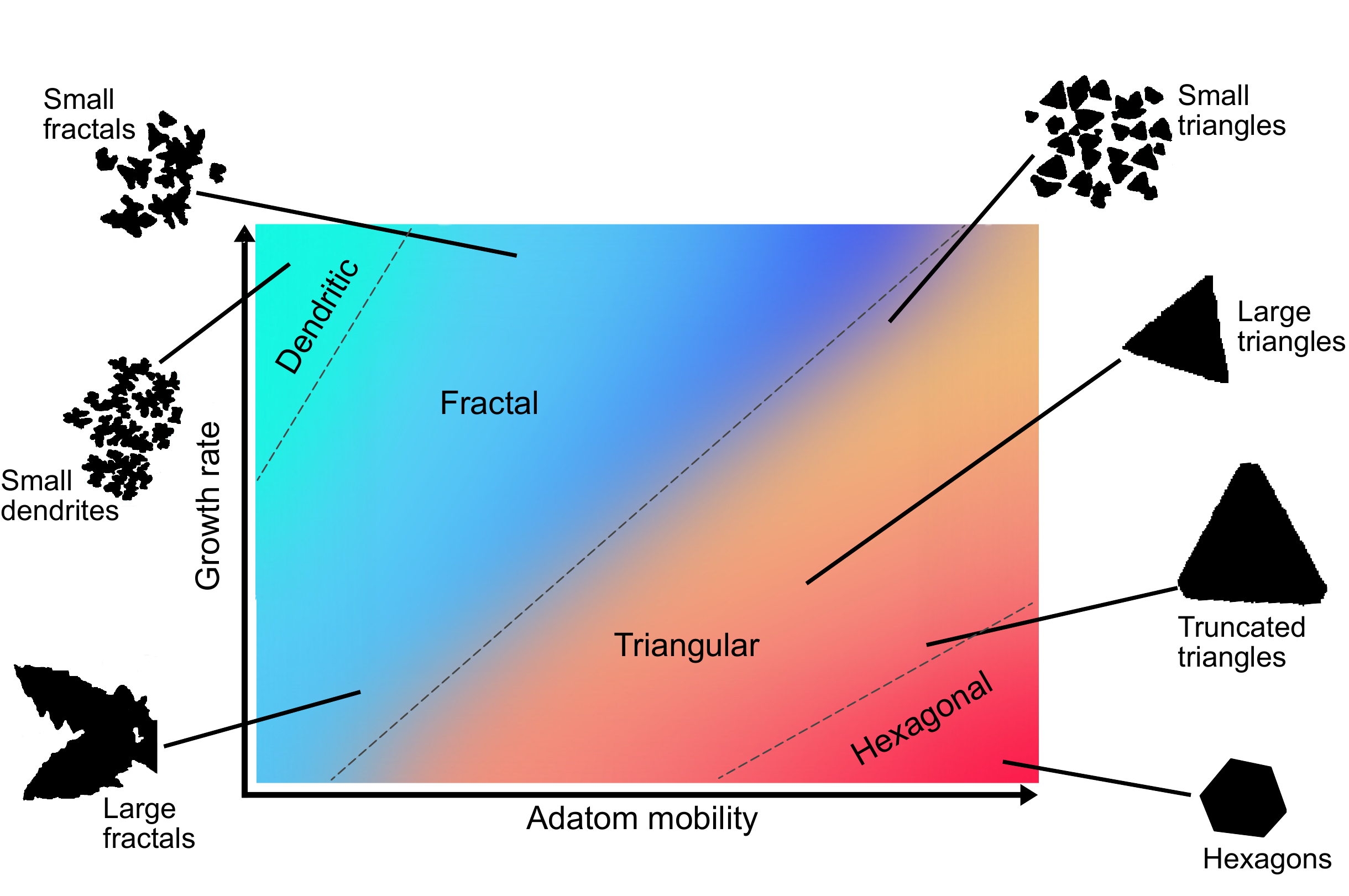}
\caption{Schematic phase diagram showing the island morphologies that can be obtained via molecular-beam epitaxy growth of monolayer transition-metal dichalcogenides, as determined from the studies presented above. The inserts show examples of the corresponding domain morphologies extracted from AFM scans. }
\label{Fig6}
\end{figure*} 
We summarise our key findings in the schematic phase diagram shown in Figure~\ref{Fig6}.  For low adatom mobilities, an undesirable dendritic growth mode is found. Within the parameter range investigated here, this was only observed for NbSe$_2$, pointing to the additional challenges for TMD growth associated with the low diffusion lengths of the heavier transition-metals, which have lower vapour pressures as compared to the lighter transition-metals. Nonetheless, morphological control for such systems is still possible. With reducing growth rate or increasing surface diffusion lengths promoted via increased growth temperature, or higher intrinsic adatom mobility of different transition metals, the dendritic growth mode transforms into a fractal mode, with tree-like branching morphologies. Within the fractal growth region, there is a clear dependence of the size of the monolayer islands on both growth rate and adatom mobility; smaller fractals are obtained when materials with smaller adatom mobilities are grown at lower temperatures under faster growth rates. The fractals get larger with an increasing adatom mobility and with reducing growth rate. 

Upon further increasing the adatom diffusion lengths and lowering the growth rate, a more thermodynamically favourable compact triangular domain growth regime can be achieved. The growth conditions in this region further promotes the transformation of neighbouring fractal domains into single triangular islands. Finally a regime where the growth of the most stable and thermodynamically favourable hexagonal domains is be obtained can be found for the highest adatom mobilities. A clear and steady transition region is observed between the triangular and hexagonal growth regimes.

Ultimately, our study therefore indicates that, to achieve large monolayer triangular or hexagonal domains, growth should proceed at high substrate temperature to promote surface adatom mobility, and at low growth rate to increase time available for surface diffusion. The high growth temperatures in turn necessitates the use of very high Se overpressures, to compensate surface desorption due to the extremely high vapour pressure of this element. The required growth conditions will vary for a given transition-metal atom used: for NbSe$_2$, growth at an extremely slow rate of ca.\ 0.05 ML/h was required to obtain domain sizes of ca.\ 150~nm for a $\sim\!0.5$~ML coverage, while for TiSe$_2$, island sizes of over 1~$\mu$m$^2$ could be achieved for a similar surface coverage at a much faster growth rate of ca. 0.5 ML/h. For TiSe$_2$, via use of the optimised growth conditions as determined here, we were able to achieve the largest monolayer islands of a TMD grown by molecular-beam epitaxy to date. 

While we studied three specific TMDs here, our conclusions should be generally applicable to the growth of other TMDs using this method.  Our study thus paves the way to the synthesis of improved-quality epilayers in challenging systems such as the $4d$ and $5d$ systems, which are of interest, for example, for their optoelectronic properties, strong spin-orbit interactions, and possibilities to stabilise exoitc quantum states.\cite{Chhowalla2013, xu2014spin, law20171t} Moreover, by further extending the parameter range studied here, our results suggest the route to even larger island sizes of the lighter $3d$ systems, which may consequently be able to approach the grain sizes achieved in other monolayer preparation methods such as mechanical exfoliation.

 \section{Experimental Section}
Materials were grown on HOPG substrates using a DCA R450 MBE system. The growth chamber has a base pressure of $\sim$ 1 $\times$ 10$^{-10}$ mbar and a background pressure of $\sim$ 3 $\times$ 10$^{-9}$ mbar during growth. HOPG substrates were chosen for the growth due to their similar crystal symmetry to the TMD epilayer, weak van-der-Waal's interactions between the substrate surface and epilayer and the thermal stability of HOPG at the highest growth temperatures used for this work. Fresh HOPG surfaces were exfoliated in atmosphere before rapidly transferring into a vacuum load lock. Substrates are first degassed at $\sim$ 200 $^\circ$C in the load lock overnight before transferring to the growth chamber. The quality of the substrate surface was monitored using $in-situ$ reflection high energy electron diffraction (RHEED). Prior to growth, the substrate is further annealed at 600 - 950 $^\circ$C for $\sim$ 20 minutes before cooling to growth temperature, which varied from 300 to 900 $^\circ$C. 

For transition metal sources, high temperature effusion cells containing 4N pure V, 3N5 pure Ti and an electron-beam evaporator containing 3N5 pure Nb were used. A valved cracker cell was used to generate 5N pure Se flux. The cracker zone of the Se source was maintained at an elevated temperature of 500 $^\circ$C during growth, to generate cracked Se monomers or dimers and to prevent condensation near the valve. During a typical growth, V and Ti fluxes were maintained at $\sim$ 6 $\times$ 10$^{-10}$~mbar beam-equivalent pressure (BEP), which was measured by positioning a retractable beam flux monitoring ion gauge in front of the substrate, just before growth. The Nb flux was measured using a flux monitor built into the e-beam assembly. A Nb flux of 1.5~nA is used for typical growths, unless otherwise specified. For this study we used a varying Se BEP from $\sim$ 1 $\times$ 10$^{-8}$ to $\sim$ 3 $\times$ 10$^{-7}$ mbar. During growth, the sample surfaces were monitored using the RHEED operated at 15~keV. 

Surface morphology analysis was performed after removing the as-grown sample from vacuum, in atmospheric conditions. A typical sample is exposed to air for $\sim$ 2-4 hours before being scanned. A Bruker Multimode atomic force microscope (AFM) was used to examine the morphology of the epilayers. Samples were scanned in tapping mode using a Si tip. The step height obtained using AFM for the monolayers studied here was in the range of $\sim$ 7.5 $\AA$, which is in very good agreement with the height value of typical transition metal dichalcogenides, where a monolayer is composed of three layers of atoms.

\section*{Acknowledgments}

We thank David Jones, Chao Dun Tan, and Georg Haehner for access to the AFM system (funded via an EPSRC equipment grant: EP/L017008/1) used in this work, and for providing experimental support. We gratefully acknowledge support from The Leverhulme Trust (Grant no. RL-2016-006), The Royal Society, and the European Research Council (Grant No. ERC-714193-QUESTDO). K.U. acknowledges EPSRC for PhD studentship support through grant no. EP/L015110/1.

\end{document}

\title{Supplementary Figure: Morphology control of epitaxial monolayer transition metal dichalcogenides}

\author{A.~Rajan}
\email{ar289@st-andrews.ac.uk}
\affiliation{SUPA, School of Physics and Astronomy, University of St. Andrews, St. Andrews KY16 9SS, United Kingdom}
\affiliation{These authors contributed equally to this work}
\author{K. Underwood}
\affiliation{SUPA, School of Physics and Astronomy, University of St. Andrews, St. Andrews KY16 9SS, United Kingdom}
\affiliation{These authors contributed equally to this work}
\author{F. Mazzola}
\author{P.D.C. King}
\email{philip.king@st-andrews.ac.uk}
\affiliation{SUPA, School of Physics and Astronomy, University of St. Andrews, St. Andrews KY16 9SS, United Kingdom}

\maketitle

\begin{figure*}[h!]
\centering
\includegraphics[width=1\textwidth]{ARPESSupplementaryV2}
\caption{Angle resolved photoemission spectroscopy (ARPES) measurements for monolayer NbSe$_2$ films grown at (a) 300$^{\circ}$C, (b) 500$^{\circ}$C and (c) 600$^{\circ}$C. For growth temperatures below 500$^{\circ}$C we observe a metallic band crossing the Fermi level. This is indicative of the metallic 1H polymorph seen in NbSe$_2$ \cite{MottNbSe2}. However, upon increasing the growth temperature to 600$^{\circ}$C we find additional spectral signatures of a gapped state $\approx$250~meV below the Fermi level. This is indicative of the 1T poymorph of NbSe$_2$ \cite{MottNbSe2}, indicating that growth at this temperature leads to a mixed phase sample comprising of both 1H and 1T polymorphs within our probing area of $\approx\!800$~$\mu$m. }
\label{S1}
\end{figure*}

%\section*{References}